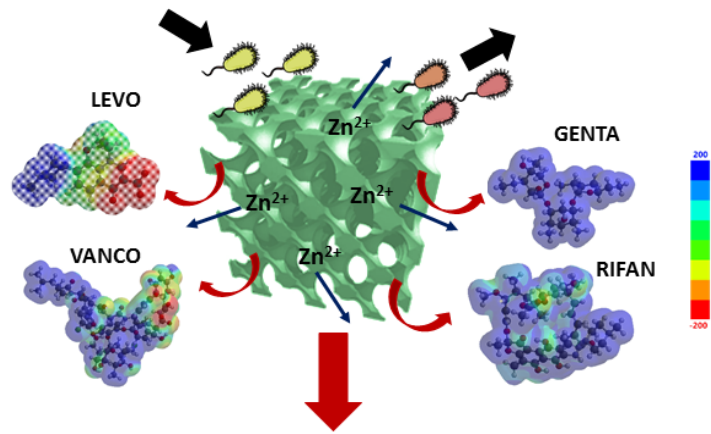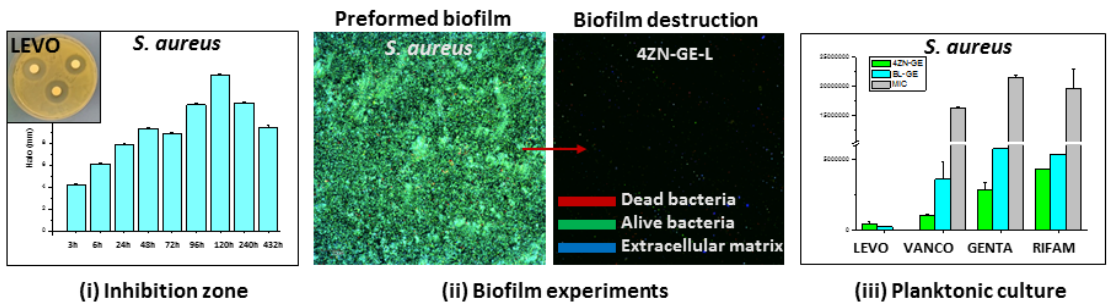

**Graphical abstract**

# Multifunctional antibiotic and zinc containing mesoporous bioactive glass scaffolds to fight against bone infection

C. Heras[a], J. Jiménez Holguín[a], A. L. Doadrio[ab], M. Vallet-Regí[ab], S. Sánchez-Salcedo[ab*], A. J. Salinas[ab]

[a] *Departamento de Química en Ciencias Farmacéuticas, Facultad de Farmacia, Universidad Complutense de Madrid, UCM, Instituto de Investigación Hospital 12 de Octubre, i+12, 28040 Madrid, Spain.*
[b] *Networking Research Center on Bioengineering, Biomaterials and Nanomedicine (CIBER-BBN), Spain.*
salinas@ucm.es, sansanch@ucm.es

**Abstract**

Bone regeneration is a clinical challenge which requires multiple approaches. Sometimes, it also includes the development of new osteogenic and antibacterial biomaterials to treat the emergence of possible infection processes arising from surgery. This study evaluates the antibacterial properties of gelatin-coated meso-macroporous scaffolds based on the bioactive glass 80%$SiO_2$–15%CaO–5%$P_2O_5$ (mol-%) before (BL-GE) and after being doped with 4% of ZnO (4ZN-GE) and loaded with both saturated and the minimal inhibitory concentrations of one of the antibiotics: levofloxacin (LEVO), vancomycin (VANCO), rifampicin (RIFAM) or gentamicin (GENTA). After physical-chemical characterization of materials, release studies of inorganic ions and antibiotics from the scaffolds were carried out. Moreover, molecular modelling allowed determining the electrostatic potential density maps and the hydrogen bonds of antibiotics and the glass matrix. Antibacterial in vitro studies (in planktonic, inhibition halos and biofilm destruction) with S. aureus and E. coli as bacteria models showed a synergistic effect of zinc ions and antibiotics. The effect was especially noticeable in planktonic cultures of S. aureus with 4ZN-GE scaffolds loaded with VANCO, LEVO or RIFAM and in E. coli cultures with LEVO or GENTA. Moreover, S. aureus biofilms were completely destroyed by 4ZN-GE scaffolds loaded with VANCO, LEVO or RIFAM and the E. coli biofilm total destruction was accomplished with 4ZN-GE scaffolds loaded with GENTA or LEVO. This approach could be an important step in the fight against microbial resistance and provide greatly needed options for bone infection treatment.

**Statement of significance**

Mesoporous bioactive glasses (MBGs) are bioceramics with unique properties which make them excellent materials for bone tissue engineering. MBGs were doped with zinc, as osteogenic and antibacterial ion, at percentages of 0%, named as BL and used as control, and 4%, named as 4ZN. Tridimensional meso-macro porous scaffolds were built up by rapid prototyping and then loaded with different antibiotics (Levofloxacin, Vancomycin, Gentamicin and Rifampicin). Those materials were physico-chemical characterized and then, tested *in vitro* with two different bacteria strains used as a model of infection (*Staphylococcus aureus* and *Escherichia coli*). The zinc-antibiotic scaffold system was tested in inhibition zone assays, planktonic growth assays and biofilm assays. Results showed a synergetic effect between zinc and the different antibiotics, obtaining great inhibition results in all three types of assays. Further, it was possible to reduce the antibiotic concentration to the Minimal Inhibitory Concentration (MIC 50%) of each antibiotic for each strain, which in combination with zinc, was able to inhibit the bacterial planktonic growth. To conclude, MBGs scaffolds enriched with zinc and loaded with the different antibiotics, constitute a great system to fight bacterial infection in its three states of growth, being able to reduce the antibiotic concentration to the MIC, which in combination with the effect of zinc, also inhibits the bacterial growth. It constitutes a great multifactorial tool in tissue engineering because its capacity to generate support to the damaged bone tissue, its osteogenic properties and its capability to fight bacterial infections with lower antibiotic dosage.

# 1. Introduction

The emergence of infection processes is one of the most common and challenging complication during and after a surgical intervention which threaten millions of lives worldwide [1,2]. Bone bacterial infection is a serious issue with devastating clinical and socio-economic implications. It is an inflammatory process which provokes bone destruction (osteolysis) [3,4]. Post-operative implant infections are one of the most serious complications associated with bone diseases and fractures, treated with bone grafts and prostheses [5,6]. Bacteria of principal concern include *Staphylococcus aureus*, *Staphylococcus epidermidis* and *Escherichia coli* [5,7,8] Approaches usually involve surgery, antibiotic systemic administration and implant removal which has deep limitations and important repercussions in patients live quality: additional surgical interventions, prolonged hospital stays, high side effects and higher mortality [9,10]. The appearance of bacterial infection derived from surgical procedures (332000 total hip and 719000 total knee arthroplasties performed in 2010, just in the United States) [11] represent a substantial problem. The numbers are projected to reach 572000 and 3.48 million by 2030 for hips and knees, respectively [11]. In Europe, the situation is analogous but larger number of patients undergoes primary hip arthroplasty than knee arthroplasty. In addition to hip and knee replacement, shoulder, elbow, and ankle arthroplasties are now available. The total number of patients with existing arthroplasties continues to increase. Modelling data predicts the incidence of prosthetic joint infection will increase to greater than 6% by 2030 [11]. owing to factors such as increased demand for surgery, the aging population and the obesity epidemic [12-14]. Moreover, the number of people over 50 years suffering from bone related diseases, was already envisaged to increase two fold from 2012 to 2020 [15]. One of the principal reasons of the failure is the appearance of antibiotic-resistant strains [16-19] and the bacteria capability to develop a biofilm. Biofilms are groups of bacteria growing together which are able to produce an extracellular matrix accomplishing in a cooperative manner [16-19]. In a biofilm, bacteria can grow protected from environmental stresses, develop antibiotic resistance and skip the innate and adaptive immune system host strategies [20,21]. Being aware that oral antibiotic administration and antibiotic injection involve several problems, higher antibiotic concentrations are required which is not a recommendable approach. Moreover, the administration of antibiotics below the Minimal Inhibitory Concentration (MIC) leads to the development of antibiotic-resistant species. Therefore, the development and use of local antibiotic release systems, able to eliminate or reduce the extent of bacterial attachment, growth, and biofilm formation have attracted significant attention in tissue engineering and regenerative medicine strategies [22,23].

Regarding this problem, the development of a multifunctional 3D scaffold based on Mesoporous Bioactive Glasses (MBGs) [24], doped with antibacterial ions [25] and loaded with low enough concentration of different antibiotics to avoid antibiotic resistance, would be an excellent approach [26,27]. MBGs are unique bioceramics because of their textural properties. During the last years, MBGs have gained importance because of their high ordered mesoporous structures, able to host and release different substances like antibiotics [28,29]. Their huge surface area and pore volume [30]. which provides fast in vitro and in vivo responses, bone regenerative capability [31]. and the possibility of being doped with different ions to add new capabilities to the biomaterial [32-34]. MBG bioceramics allow local delivery, which can reduce potential toxicity risk, costs and time efficacy. It also improves bioavailability, being able to minimize its concentration and avoid drug resistances. Nowadays, one of the most interesting approaches is the possibility of combining different metal ions and antibiotics in order to reduce antibiotic concentration, supplying the antibiotic effect with the therapeutic effect of specific metal ions. It has been demonstrated that the variation of intracellular ion concentration caused by the bioactive glass biodegradation, leads the activation of intracellular signalling pathways [33]. Mainly, zinc, copper and silver ions, incorporated into the glass structure, allow the release of biological active ions with antibacterial properties [35,36]. Moreover, they can influence gene expression mechanisms of osteoprogenitor cells, increasing bone regeneration [33]. Zinc ions would be a good option for that purpose, due to their specific capacity to increase osteogenesis, angiogenesis and their antimicrobial properties [37]. There are several studies describing the antibacterial effect of zinc, the use of zinc in combination with antibiotics and the use of zinc as an alternative therapy to fight infection. However, there is no evidence of scaffolds, which acts as a biocompatible, temporary and osteogenic matrix for bone proliferation, doped with zinc as osteogenic, angiogenic and antibacterial ion, and loaded with different antibiotics at their Minimal Inhibitory Concentration (MIC) to fight different infection processes: Inhibition zone, planktonic growth and biofilm development and destruction.

In this context, the aim of the present study was to investigate the antibacterial effect of zinc-enriched MBG scaffolds loaded with antibiotic. MBG (80%$SiO_2$– 15%CaO–5%$P_2O_5$ (mol-%) scaffolds (BL) and an analogous doped with 4% of ZnO (4ZN), coated with gelatin (BL-GE and 4ZN-GE) and loaded with one of the next antibiotics Levofloxacin (LEVO), Vancomycin (VANCO), Gentamicin (GENTA) or Rifampicin (RIFAM) (BL-GE-L/V/G/R and 4ZN-GE-L/V/G/R) were tested in *S. aureus* and *E. coli* strains as bone infection bacteria models. Because a single antibiotic will be loaded each time, it will be possible to study the interactions of each one with the scaffold, its anti-bacterial capacity against the model strains proposed and the potentiating effect for each antibiotic of acting in conjunction with zinc ions released from the scaffold.

## 2. Materials and methods

*2.1. Reactants*

Reactants were purchased from Sigma-Aldrich (St. Louis, USA) Inc. to perform the described experiment: tetraethyl orthosilicate (TEOS), triethyl phosphate (TEP), calcium nitrate tetrahydrate $(CaNO_3)_2 \cdot 4H_2O$), pluronic F123 (EO20PO70EO20), ε-caprolactone (Mw = 58,000 Da) and hydrochloric acid (HCl, 37 %). Absolute ethanol (EtOH) was purchased from Panreac Quimica SLU (Castellar del Valles, Barcelona, Spain). All reagents were used as received without further purification. Ultrapure deionized water with resistivity of 18.2 MΩ was obtained using a Millipore Milli- Q plus system (Millipore S.A.S., Molsheim, France).

*2.2. Synthesis of mesoporous bioactive glass scaffolds*

The production of scaffolds with compositions of 80%$SiO_2$–15%CaO–5%$P_2O_5$ (mol %) (BL scaffolds) and an analogous including 4% ZnO, 82.2%$SiO_2$–10.3%CaO–3.3%$P_2O_5$–4.2%ZnO (mol %) (4ZN scaffolds) with sizes of 7 mm diameter and 10 mm height, was described in detail by Heras et al [38]. As in the previous publication, scaffolds were coated with a thin layer of gelatin crosslinked with glutaraldehyde (GA) and named as BL-GE and 4ZN-GE corresponding to the zinc undoped scaffolds and the 4%ZnO doped scaffolds, respectively. The previous study used, shorter scaffolds (7 mm diameter and 5 mm height) but all their physical chemical properties were identical than used in the present study.

*2.3 Meso-macroporous 3D scaffolds characterization*

Scaffolds were characterized by transmission electron microscopy (TEM), scanning electron microscopy (SEM), energy dispersive X-ray (EDX) spectroscopy, Fourier transform infrared FTIR spectroscopy and small angle X-ray diffraction (SA-XRD) as described in ref. 38. Moreover, thermogravimetric analysis (TGA) and Nitrogen adsorption measurements were done in the raw scaffolds, as described in ref. 38, and in the drug loaded scaffolds to verify the antibiotic loading process.

*2.4 In vitro antibiotic adsorption and release*

Scaffolds were loaded with each antibiotic by immersion. After sterilizing with UV radiation (30 min each side), scaffolds were placed in a 12 well plate and dissolving the desired antibiotic amount in 3 mL of the adequate solvent for each one. Depending on the experiment, it was used saturated antibiotic concentration or minimal inhibitory concentration (MIC = 50%) of antibiotic for each strain. Immersed scaffolds were kept 24 h at room temperature under stirring (100 rpm). Then, scaffolds were vigorously washed and dried 48 h at 37 °C. It is important to highlight that LEVO and RIFAM are light sensitive, so the entire process was done in dark conditions to avoid light degradation. Dry scaffolds were immersed in phosphate buffered saline (PBS) solution adjusted at physiological pH of 7.4. Pieces were kept at 37 °C in stirring conditions (100 rpm) up to 18 days. 500 µL aliquots were extracted each day to measure the antibiotic release by UV spectrophotometry. The whole antibiotic release procedure was also performed at pH 6.5, which corresponds with the acidic pH generated in an infection process. There were no significant differences between both pH and because of that release results at pH 6.5 are not shown.

*2.5 Molecular modelling*

A molecular model based on mesoporous MCM-41 silica doped with zinc, similar to mesoporous $SiO_2$–$CaO$–$P_2O_5$–$ZnO$ glass, was generated by using Spartan´14 software (Wavefunction Inc.). The building unit in this model was the $Si_6O_{12}$ pseudo–cell, consisting on hexagonal arrangements of Si–O–Si and Si–O–Zn units until reaching a model of 2 nm length and 4 nm diameter. The final structure was refined by geometry optimization using the MM+ force field. Docking calculations were carried out on the MCM-41 surface model by Hex Cuda 8.0 software with a cubic box of 10 nm and grid dimension of 0.6 nm. The correlation was only shape type. For electrostatic potential maps, coordinates of gentamicin 5+, zwitterionic rifampicin, zwitterionic levofloxacin and vancomycin 1+ were obtained from PubChem (U.S. National Library of Medicine) and treated by molecular dynamics simulations from Spartan´14 (Wavefunction, Inc.).

*2.6 Ion release assay*

For these assays, each scaffold was placed in a well of a transwell plate of 12 wells with 2 mL of Todd Hewitt Broth (THB) in each of them. They were kept at

37 °C with stirring (100 rpm) for 10 days. Each day, the 2 mL were extracted in full to measure in the culture medium the amount of Ca, P and Zn ions that had been released and another 2 mL of fresh THB were added to continue with the release study. Ion concentrations were determined by inductively coupled plasma/optical emission spectrometry (ICP/OES) using an OPTIMA 3300 DV device (Perkin Elmer). Each ion was determined in two different samples measured by triplicate [38].

*2.7 Bacterial cultures*

*2.7.1 Agar disk-diffusion tests of drug loaded materials*

Zinc-doped scaffolds (4ZN-GE scaffolds) and loaded with the saturated concentration of each antibiotic, were immersed in 2 mL at pH 7.4 and 6.5 and kept at 37 °C and stirring conditions (100 rpm). 7.4, was selected because corresponds to physiological one, and 6.5 because matches with generated in infection environment. Aliquots (20 µL) were extracted at several times to load foams placed later in contact with a pre-seeded (100 µL) of *S. aureus* or *E. coli* having $2 \times 10^6$ bacteria/mL) agar plates. After 24 h of culture in static conditions at 37 °C, the inhibition zone was analysed. The bacterial inhibition zone size was measured as: (outer diameter of the inhibition zone - disk diameter)/2. Each study was performed in triplicate.

*2.7.2 Planktonic growth inhibition test*

Pre-loaded scaffolds, at saturated or minimal inhibitory concentration (MIC) of each antibiotic for each strain (*S. aureus* and *E. coli*), were placed in a 12 transwell plate. Then, 3 mL of bacteria dissolution ($2 \times 10^8$ bacteria/mL) were added to put in contact bacteria culture with the zinc doped (4ZN-GE) or undoped (BL-GE) scaffold loaded with the different antibiotics. 24 h later, aliquots (20 µL) were extracted and seeded in agar plates. After 24 h of culture at 37 °C in static conditions, colony forming units (CFUs) were counted to know how the material, ions and antibiotic, affect the bacterial growth. Results of 4ZN-GE-antibiotic loaded scaffolds and BL-GE-antibiotic loaded scaffolds were compared. For the tests at saturated drug concentration, bacteria culture controls without material and controls with bacteria cultured with drug unloaded materials were carried out. For MIC tests, controls were carried out with bacteria cultures, growing in contact with the MIC of each drug for each strain, but without any material.

*2.7.3 Biofilm degradation*

1 mL of both strains (*S. aureus* and *E. coli*) was seeded ($2x10^8$ bacteria/mL) on a glass slide, kept at 37 °C and stirring conditions (100 rpm). During the first 24 h of culture, the medium (THB) was doped with 4% of sucrose. Then, it was replaced by a normal one. 24 h later, the biofilm was already formed. Pre-formed biofilms were placed in a 12 transwell plate were the different type of scaffolds were also located (BL-GE, 4ZN-GE and 4ZN-GE-Drug loaded scaffolds). After 24 h of culture at 37 °C and stirring conditions (100 rpm), the biofilm was extracted and stained with live and dead BacLigth bacteria viability kit (Thermo Fisher Scientific, Invitrogen[TM]) and tested in a MC1025 confocal laser scanning microscope (Biorad). Alive bacteria were stained in green (SYTO 9), dead bacteria in red (propidium iodide, PI) and extracellular matrix in blue (calcofluor).

## 3. Results and discussion

### 3.1. Meso-macroporous 3D scaffolds characterization

Structural powder characterization by SA-XRD in Fig. 1, revealed that gelatin containing scaffolds, exhibited maxima in the mesoporous order indicative region. However, differences between samples as a function of the ZnO content were observed. Thus, BL-GE displayed a sharp diffraction maximum at 2θ in the region of 1.0–1.4°, assigned to the (10) reflection along with a poorly resolved peaks at around 2.0 that can be assigned to the (11) reflection. These maxima were indexed on the basis of an ordered two-dimensional (2D) hexagonal structure (plane group p6mm). The intensity of the (10) maximum decreased while the ZnO content increased, indicating a partial deterioration of the mesoporous structure in 4ZN-GE scaffolds. TEM images of zinc-doped (4ZN) and undoped (BL) scaffolds, obtained with the electron beam parallel to the mesoporous channels, are shown in Fig. 1. All of the samples exhibited a typical 2D-hexagonal ordered mesoporous arrangement. FITR profile of the 4ZN, 4ZN-GE, BL and BL-GE scaffolds showed the characteristic bands of each material, including the bands of gelatin crosslinked with GA used to coat the scaffolds (Fig. 2A). SEM micrographs from 4ZN-GE scaffolds, showed two types of designed channels (Fig. 2B-D). First of them, with sizes between 700 and 1000 µm, and the second ones, between 1000 and 1500 µm (Fig. 2B). At higher magnification, it was possible to appreciate in both, surface and fracture views of the 300 µm wide bars, of macropores between 1-10 µm and to visualize the homogeneous gelatin layer on the material surface (Fig. 2C-D). BL-GE material showed similar SEM structure to 4ZN-GE. Therefore, BL-GE images were omitted for Fig. 2 simplification.

*3.2. Ion release assay*

To understand the possible antibacterial activity of the scaffolds, the cumulative release of calcium, phosphorus and zinc ions after soaking in THB was measured (Fig. 3). For BL-GE scaffolds, the lower release of $Ca^{2+}$ ions compared with 4ZN-GE was attributed to its higher in vitro bioactive behaviour. Thus, a part of the $Ca^{2+}$ ions released would precipitate as carbonate hydroxyapatite (CHA) on the scaffold surface being eliminated from medium. In contrast, 4ZN-GE scaffolds displayed slower bioactive behaviour and, consequently, most part of the released $Ca^{2+}$ ions from these scaffolds remained in the solution. Regarding phosphorous ions, 4ZN-GE scaffolds showed lower concentration in the medium that was attributed to the precipitation of calcium and zinc phosphates [28,38]. Accumulated concentration of zinc increased about 1.5 ppm/day in the first 48 h and more slowly for longer times, in the same conditions than those of bacterial assays.

*3.3. Antibiotic loading in MBG scaffolds*

Antibiotic adsorption was performed at two different concentrations: saturated and the Minimal Inhibitory Concentration (MIC) for each antibiotic and tested strain.

*3.3.1. Thermogravimetric Analysis*

Thermogravimetric analysis (TGA) of scaffolds provided the amount of drug loaded in each case (Table 1). This analysis showed the satisfactory drug loading into the scaffolds by detecting losses of mass which correspond to the loaded drug.

*3.3.2. Nitrogen Adsorption Porosimetry*

Nitrogen adsorption–desorption isotherms and pore size distributions of BL-GE and 4ZN-GE scaffolds, before and after loaded with each antibiotic, are shown in Fig. 4. All curves can be identified as type IV isotherms, characteristic of mesoporous materials (Fig. 4.A). Observed type H1 hysteresis loops in the mesopore range are characteristic of open cylindrical pores. Fig. 4.B corresponds to pore size distributions. Variations were detected when the scaffolds were loaded with the antibiotics. Loaded scaffolds showed a decrease both in surface area and pore volume verifying the drug loading into the mesopore structure. Table 1 collects the textural properties, i. e., surface area ($S_{BET}$), pore diameter

($D_P$) and pore volume ($V_p$) of loaded and non-loaded scaffolds. In general, higher values were found for antibiotic unloaded samples. Thus, BL-GE and 4ZN-GE antibiotic-loaded scaffolds, presented a slight decrease in surface area and pore volume because of the drug loading. These data allowed to verify the successful drug loading into the mesopores of MBG scaffolds.

*3.4. Molecular modelling*

Mesoporous bioactive glasses (MBGs) have compositions based in $SiO_2$–$CaO$–$P_2O_5$ (BL) or $SiO_2$–$CaO$–$P_2O_5$–$ZnO$ (4ZN) systems. There, silica and phosphorus elements, as $SiO_2$ and $P_2O_5$, are known to be network formers: tetrahedral units bonded through covalent bonds by their oxygen atoms. Calcium, as CaO, is known to be a network modifier bonded through ionic interactions to the oxygen atoms. Our research group demonstrated from $^{29}Si$ NMR spectroscopy measurements that 4ZN-GE scaffolds contain [$ZnO_4$] tetrahedra. These tetrahedra present a negative charge (2–), which justifies the attraction of $Ca^{2+}$ and $Zn^{2+}$ ions acting as network compensators of charge and not as network modifier cations. Accordingly, the number of non bonding oxygen (NBO) decreases with the increase of $Q^4$ species and the decrease of $Q^3$ species [32]. Moreover, these MGBs present ionized silanol Si–OH groups in their surface which produce a negative charge that interact with the loaded drugs.
Electrostatic potential mapped density (Fig 5.A) and molecular modelling interaction studies (Fig 5.B) of 4ZN-GE material loaded with LEVO VANCO, GENTA, and RIFAM were calculated with Hex 8.0 software. Molecular modelling indicates a positive total charge in VANCO and GENTA and a zwitterionic nature in RIFAM and LEVO. Moreover, modelling showed strong 6 and 4 hydrogen bonds with VANCO (-295.7 kcal/mol) and GENTA (-250.1 kcal/mol), respectively. In case of RIFAM (-226.1 kcal/mol) and LEVO (-368.7 kcal/mol) modelling exhibited 2 and 0 hydrogen bonds, respectively. Although VANCO has 6 hydrogen bonds that favourably interact with 4ZN-GE surface, it is a large molecule (1450.3 uma) which causes low antibiotic loading. Zwitterionic LEVO presents no hydrogen bonds but it is the smallest molecule (823.9 uma) allowing the formation of a stable complex and achieved a high loading yield.

*3.5. Antibiotic release from MBG scaffolds*

Antibiotics release assays were performed in PBS at physiological pH of 7.4. Fig. 6 shows the release from 4ZN-GE scaffolds of each antibiotic along time.

Antibiotics release kinetics from the MBGs were evaluated according to zero-order kinetics and first-order kinetic model. Zero-order kinetics,

$$Q_0 - Q_t = k_0 \cdot t \quad (1)$$

Where $Q_t$ is the amount of drug remaining as a solid at time t, $Q_0$ is the initial amount of drug in the pharmaceutical dosage form and $k_0$ is the zero-order release rate constant. As it is observed in Fig. 6, LEVO presents a release preserved in time, maintaining the release even after 15 days. It must be considered that LEVO presents a zwitterionic nature (pKs 6.1-8.2) [39] having a neutral total charge but being able to establish electrostatic interactions with the silanol groups of the MBG matrix and providing a maintained release favoured by its high load and solubility [39]. However, in case of RIFAM, VANCO and GENTA, the release data showed in Fig. 6 fit better to a first-order kinetic model with an empirical non-ideality factor (δ) [40].

$$Y = A(1-e^{k_1 t})^\delta \quad (2)$$

Where Y is the percentage of antibiotic released at time t, A the maximum amount of antibiotic released (in %), and $k_1$, the release rate constant. δ values are 1 for first-order kinetics materials and 0, for materials that release the loaded drug in the very initial time of analysis. RIFAM, VANCO and GENTA present zero and positive total charge of 6+ and 5+, respectively at the physiological pH of assay. δ value provides a fidelity degree approximation to proposed model for theoretical first-order kinetics (Table 2). When δ is near to 1, first-order kinetic is more accurate and drug delivery from mesopores has smaller burst effect. This effect is usually attributed to the adsorption of drug molecules into the mesopores of the matrix. As it can be observed in Table 2, the biggest value of δ is 0.83 for RIFAM and then VANCO, 0.45 and finally GENTA 0.27. As is observed in Fig. 6, RIFAM release is maintained due to its stable interaction with the surface of MBGs and its limited solubility. However, VANCO present a less stable interaction with MBGs and GENTA a higher solubility having, in both cases, a faster release.

At this pH of 7.4, interactions between the antibiotics and the thin gelatin layer coating the scaffolds were not considered, due to the pKs: pKb ~ 6.5 and pKa ~ 4.7 of $NH_2$ and COOH ionisable groups of gelatin that were not able to establish electrostatic interactions with the drugs.

On the other hand, the complete antibiotic release assay was also performed at pH 6.5, which corresponds to the pH generated in an infection process. There were no significant differences for results obtained at pH 7.4 and at pH 6.5. For this reason, the antibiotic release results at pH 6.5 are not shown.

*3.6. Agar disk-diffusion tests of drug loaded materials*

*3.6.1. S. aureus*

4ZN-GE scaffolds were loaded with LEVO, VANCO and RIFAM because their referenced effectiveness against *S. aureus* (Fig. 7.1-3) [41-43]. The antibiotic loaded scaffolds were named as 4ZN-GE-L, 4ZN-GE-V and 4ZN-GE-R, respectively. The assay allowed evaluating the capability of the scaffolds loaded samples to eradicate bacteria. Disks foams were impregnated with aliquots (20 µL) containing the antibiotic and zinc ions released from scaffolds at each time at both infection and physiological conditions, and put in contact with pre-seeded *S. aureus* agar plates. 4ZN-GE-L scaffolds inhibited *S. aureus* growth for 18 days (Fig. 7.1) and reached their maximal inhibition after 5 days with a 14 mm halo. 4ZN-GE-V scaffolds inhibited *S. aureus* growth for 5 days. 4ZN-GE-V scaffolds reached their maximal inhibition after 6 h showing 18 % of inhibition respect LEVO (Fig. 7.2). 4ZN-GE-R scaffolds inhibited *S. aureus* growth for 10 days and reached their maximal inhibition after 48 h showing 57 % of inhibition compared to LEVO (Fig. 7.3). There were no significant differences between results obtained at physiological, 7.4, and infection, 6.5, pH values.

*3.6.2. E. coli*

4ZN-GE scaffolds were loaded with LEVO, GENTA and RIFAM active antibiotics against *E. coli* (Fig. 7.4-6) [41,44,45]. Loaded scaffolds were named as 4ZN-GE-L, 4ZN-GE-G and 4ZN-GE-R, respectively. Agar disk diffusion tests were performed as above. 4ZN-GE-L scaffolds inhibited *E. coli* growth for 18 days (Fig. 7.4). 4ZN-GE-L scaffolds reached their maximal inhibition after 72 h with a 15 mm halo. 4ZN-GE-G scaffolds inhibited *E. coli* growth for 72 h with a maximal inhibition at 6 h, showing 20% of inhibition respect LEVO (Fig. 7.5). 4ZN-GE-R scaffolds inhibited *E. coli* growth for 120 min and reached their maximal inhibition after 48 h showing 4 % of inhibition respect LEVO (Fig. 7.6). As Fig. 7 shows, the most effective antibiotics against *E. coli* were LEVO and GENTA. The inhibition zone created by RIFAM was just the 4% respect LEVO. There were no significant differences between physiological and infection pH values.

*3.7. Planktonic growth inhibition tests*

Tested scaffolds were loaded with a saturated antibiotic concentration and the loaded drug was quantified (Table 1) to assay the antibiotic effect against bacteria. The antimicrobial effect of BL-GE and 4ZN-GE scaffolds, with and without drugs, at concentrations of 27 and 80 mg scaffold/mL, was evaluated between 2 and 48 h by measuring colony forming units (CFUs)/mL. Fig. 8 shows the CFUs/mL for both strains (*S. aureus* and *E. coli*) and both materials (BL-GE and 4ZN-GE), loaded or not with the tested antibiotics. Regarding *S. aureus* cultures, bacteria growth for VANCO, LEVO and RIFAM loaded BL-GE and 4ZN-GE scaffolds was inhibited at a 99.9%. For unloaded 4ZN-GE scaffolds, only after 48 h the bacteria growth decreased up to a 50% compared to BL-GE scaffolds. This effect can be attributed to the zinc release, reaching 5.5 ppm of $Zn^{2+}$ at 2 days of assay, enough amounts to inhibit S. *aureus* growth (Fig. 3) [46]. In *E. coli* cultures, for unloaded BL-GE and 4ZN-GE scaffolds, bacteria growth was inhibited only the first 2 h until a 75% due to the calcium and zinc ions release. For BL-GE and 4ZN-GE LEVO and RIFAM loaded scaffolds, almost of 100% of bacteria growth inhibition was achieved. RIFAM was no tested against *E. coli* because it was reported that this antibiotic is not effective against these bacteria [47]. As Fig. 8 shows, the same antibiotics that managed to inhibit bacterial growth in the inhibition zone experiments, achieved an effective antibacterial effect for the planktonic growth of *S. aureus* and *E. coli*.

*3.8. Minimal Inhibitory Concentration (MIC) tests*

Previous planktonic growth tests showed no significant differences between zinc doped and undoped scaffolds. Antibiotics were very effective against both strains and zinc effect was no appreciable. For this reason, we considered of interest to prove, in an antibiotic resistant scenario, what role zinc could play. Thus, to test zinc antibacterial effect and its possible synergistic effect with antibiotics, the concentration of antibiotic loaded was reduced for each strain from the saturated one to the MIC of each drug. (Fig. 9)

Two different concentrations (27 mg/mL and 80 mg/mL) of BL-GE and 4ZN-GE scaffolds were loaded with the antibiotics to have the Minimal Inhibitory Concentration (MIC) of antibiotic for each strain released to the culture medium. Both bacteria strains were cultured 24 h in contact with the loaded scaffolds and its antibacterial effect was compared for both strains with the MIC effect for each antibiotic. Inhibition growth was achieved by all of them, compared with the MIC [48].

In *S. aureus* assays (Fig. 10) when concentration was 80 mg scaffold/mL, 4ZN-GE loaded with antibiotic achieved a higher CFUs decrease at 24 h (except for LEVO) than antibiotic loaded BL-GE scaffolds. This fact was attributed to the synergistic effect of zinc ions and the antibiotics. On the other hand, when the concentration was 27 mg scaffold/mL, the synergistic effect at 24 h was demonstrated with the combinations of zinc-containing scaffolds with VANCO or with LEVO.

In *E. coli* assays (Fig. 10) for 27 and 80 mg scaffold/mL concentrations, 4ZN-GE and BL-GE scaffolds loaded with LEVO showed almost total capability to kill bacteria at 24h. However, for GENTA-loaded scaffolds, a superior behaviour was achieved for 4ZN-GE scaffold indicating again a synergistic effect of zinc and GENTA.

*3.9. Biofilm degradation*

Finally, the antimicrobial effect of antibiotic loaded and unloaded BL-GE and 4ZN-GE scaffolds was investigated in a simulate infection environment containing *S. aureus* and *E. coli* biofilms. In Fig. 11, it is possible to appreciate the typical structure of a preformed biofilm which shows colonies of living bacteria (green) covered by a protective mucopolysaccharide matrix (blue) used as a control in both strains.

After contacting Gram + and - bacteria with the tested scaffolds, notable differences were observed in the biofilms. According to each case, they were partially or totally destroyed showing extracellular matrix in blue (calcofluor), alive bacteria in green (PI) and dead bacteria in red (SYTO 9).

Regarding *S. aureus* assays, the scaffolds BL-GE, 4ZN-GE and 4ZN-GE-G produced a partial destruction of the biofilm at 24 h. However, RIFAM, VANCO and LEVO loaded 4ZN-GE scaffolds reached complete biofilm destruction, observing colonial killed bacteria without the presence of protective layer of mucopolysaccharides.

Regarding *E. coli* assays, BL-GE, 4ZN-GE, 4ZN-GE-V and 4ZN-GE-R loaded scaffolds reached a partial destruction of the biofilm after 24 h. Nevertheless, LEVO and GENTA loaded 4ZN-GE scaffolds reached complete biofilm destruction.

These results show that our therapeutic ion-drug systems formed by the combination zinc together with GENTA or LEVO for *E. Coli* and RIFAM, VANCO or LEVO for *S. aureus* are very effective for the total destruction of the biofilm in the first 24 h of incubation, which is indicative of their antimicrobial effect. These results agree with inhibition zone assays (Fig. 7) and planktonic growth experiments (fig. 8) where it was possible to confirm the antibacterial efficiency of same antibiotics for same strain [41-45]. When biofilms were in contact with the unloaded 4ZN-GE scaffolds a partial biofilm destruction was found in both strains which show zinc antibacterial effect by itself.

## 4. Conclusions

$Zn^{2+}$ ions in combination with antibiotics, included in the mesoporous bioactive glass scaffolds increased the antibacterial effect of the system. In vitro studies of Zn-free (BL-GE) and Zn-containing (4ZN-GE) scaffolds with *S. aureus* and *E. coli*, showed a synergistic antibacterial effect of zinc with the antibiotics LEVO, VANCO, RIFAM or GENTA loaded in the scaffolds.

In planktonic assays, BL-GE and 4ZN-GE scaffolds loaded with saturated antibiotic concentration, exhibited total inhibition growth of *S. aureus* and *E. coli* cultures at 2 h. However, when scaffolds were loaded only with the MIC of antibiotics, zinc ions coming from 4ZN-GE scaffolds, showed synergistic antibacterial effect with LEVO and VANCO against *S. aureus* and with LEVO and GENTA against *E. coli* cultures.

In the inhibition growing zone tests, inhibition halos were achieved in *S. aureus* cultures at 18, 9 and 5 days with 4ZN-GE scaffolds loaded respectively with LEVO, RIFAM and VANCO. Nevertheless, in *E. coli* cultures, inhibition halos were maintained at 8 and 3 days with 4ZN-GE scaffolds loaded with LEVO and GENTA.

Partial biofilm destruction was reached with 4ZN-GE scaffolds, and total biofilm destruction with VANCO, LEVO and RIFAM 4ZN-GE loaded scaffolds for *S. aureus* and LEVO and GENTA 4ZN-GE loaded scaffolds for *E. coli*.

Therefore, the study demonstrated that the addition of antibacterial zinc ions inhibits planktonic bacterial growth and destroys biofilms with minimal antibiotic loaded concentration. There was an antibacterial effect of same antibiotics for same strains, potentiated by zinc. This approach reduces antibiotic resistance mechanisms eradicating bacteria in the surroundings of the implant site. These results provide significant insights for designing new bone osteogenic implants capable of playing simultaneously a dual role against infection.

**Acknowledgements**

This research was funded by Instituto de Salud Carlos III, grant number PI15/00978 co-funded with European Union FEDER funds and the European Research Council, Advanced Grant Verdi-Proposal No. 694160 (ERC-2015-AdG).

# Figure captions

**Figure. 1.** SA-XRD of the 4ZN and BL scaffolds and TEM micrographs of the zinc doped (4ZN) and undoped (BL) scaffolds.

**Figure. 2. A**. FTIR spectra of the 4ZN, 4ZN-GE, BL and BL-GE scaffolds. SEM micrographs of 4ZN-GE scaffold coated with GA crosslinked gelatin: **B.** Front view with channels of 700–1000 µm and 1000–1500 µm with layers at 45º. **C.** Detail of two cross macroporous bars of 300 µm of diameter. **D.** Fracture section of porous bars (1-10µm).

**Figure. 3**. Evolution of $Ca^{2+}$, P (V) and $Zn^{2+}$ ions released from Mesoporous Bioactive Glass (MBG) scaffolds soaked in Todd Hewitt broth (THB) as a function of time.

**Figure. 4.** Nitrogen adsorption isotherms (A, A´) and pore size distributions (B,) of BL-GE and 4ZN-GE before and after be loaded with each antibiotic.

**Figure. 5 A.** Molecular Docking from Hex 8.0 and CUDA 5.0 Nvidia GPU acceleration software with Polar fast Fourier transform (FFT) method and Shape+electro+DARS (decoys as reference state) correlation type. 4ZN-GE was used as receptor and gentamicin 5+, levofloxacin zwiterion, rifampicin zwiterion and vancomycin 1+ as ligands. H-bonds are represented in dot line. **B.** Electrostatic potential mapped density for Zn-Ca modified MCM-41 and gentamicin 5+, levofloxacin zwitterion, rifampicin zwitterion or vancomycin 1+ from Hex 8.0 software.  The energy was optimized by AM1 Hamiltonian Method.

**Figure. 6**. Antibiotic release from 4ZN-GE scaffolds at physiological pH.

**Figure. 7**. 4ZN-GE-L (**1**), 4ZN-GE-V (**2**) and 4ZN-GE-R (**3**) scaffolds inhibition zone against *S. aureus,* and 4ZN-GE-L (**4**), 4ZN-GE-G (**5**) and 4ZN-GE-R (**6**) scaffolds inhibition rate against *E. coli* along the time, at physiological pH of 7.4 and at pH of 6.5 of an infection environment. * Indicates significant differences between pH 7.4 and 6.5. Statistical significance: $p < 0.05$.

**Figure. 8**. Planktonic growth of *S. aureus* and *E. coli* after 2, 6, 24 and 48 h of bacteria culture in contact with zinc doped (4ZN) and undoped (BL) scaffolds, loaded and unloaded with different drugs: Levofloxacin (L), Vancomycin (V), Gentamicin (G) and Rifampicin (R). * indicates significant differences between 2, 6, 24 and 48 h. Statistical significance: $p < 0.05$.

**Figure. 9.** Antibiotic MIC loaded and released from 4ZN-GE scaffolds at physiological pH (7.4) after 6 and 24 h. Release of antibiotics from BL-GE was analogous for all the antibiotics (data not shown).

**Figure. 10**. CFUs/mL of *S. aureus* and *E. coli* for 27 and 80 mg scaffold/mL concentrations after 24 h in contact with 4ZN-GE and BL-GE antibiotic loaded scaffolds and with MIC. *and # indicate significant differences between the different samples. Comparisons between: BL-GE and 4ZN-GE (*), MIC and BL-GE (#); Statistical significance: $p < 0.05$.

**Figure. 11.** Confocal micrographs of *S. aureus* and *E. coli* biofilms before (control) and after 24 h in contact with BL-GE scaffolds, 4ZN-GE scaffolds and 4ZN-GE scaffolds loaded with each tested antibiotic

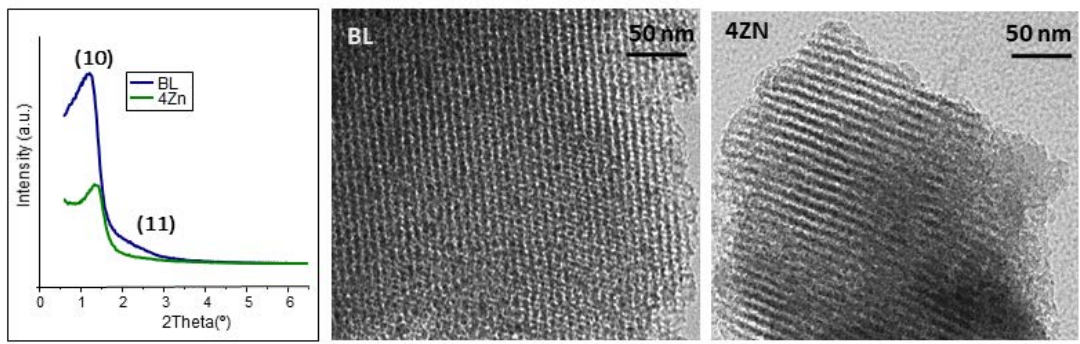

Figure 1

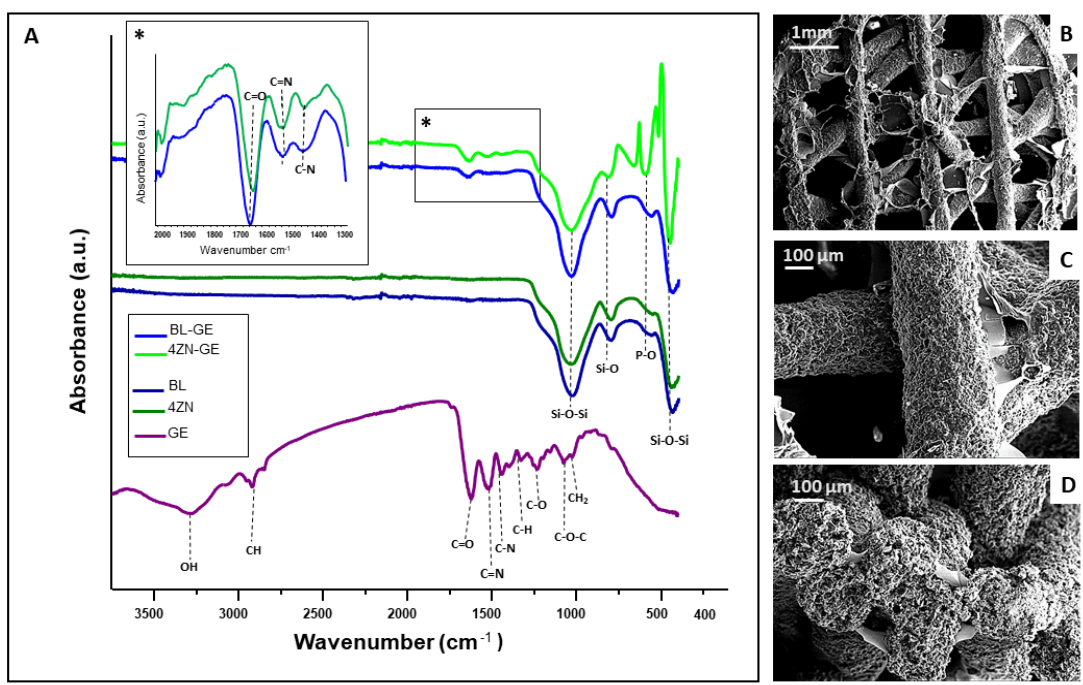

Figure 2

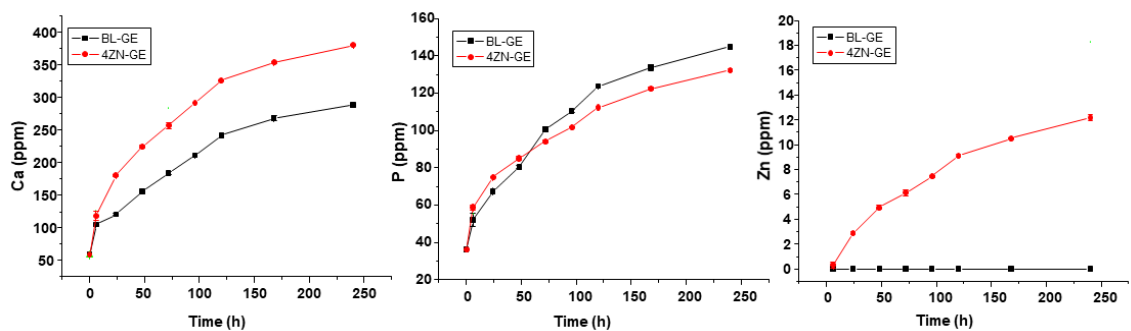

Figure 3

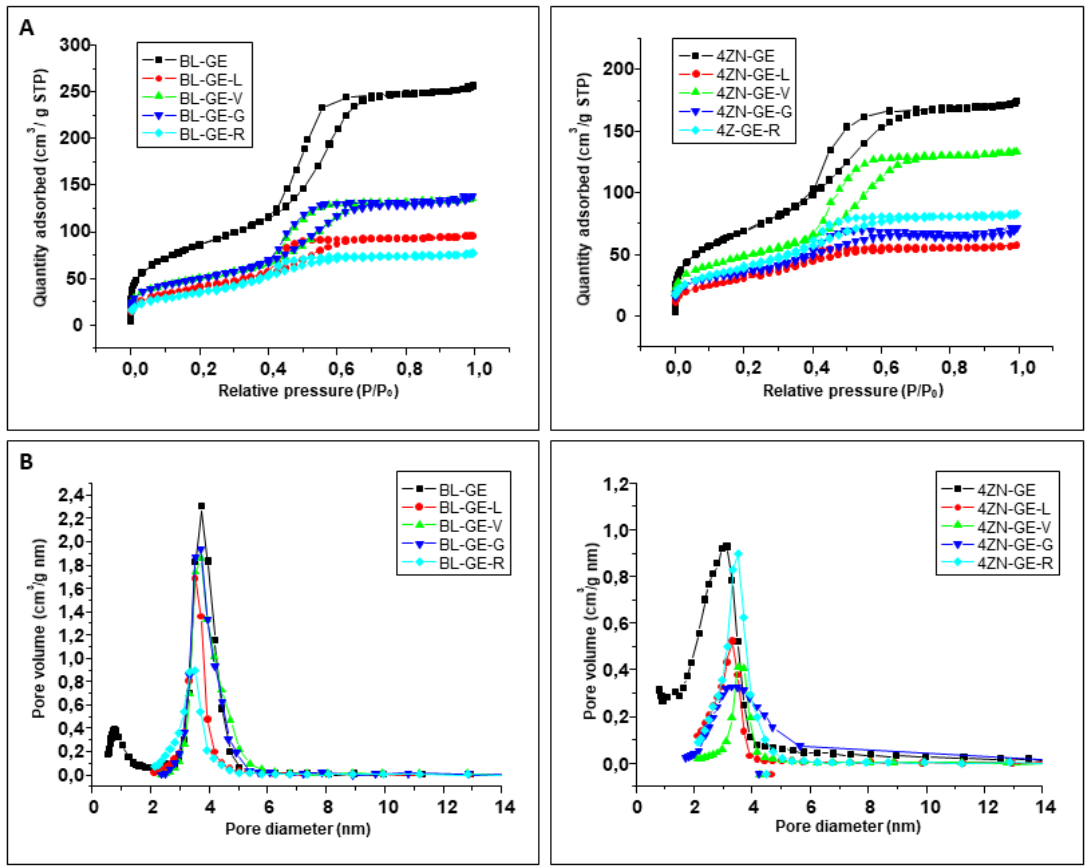

Figure 4

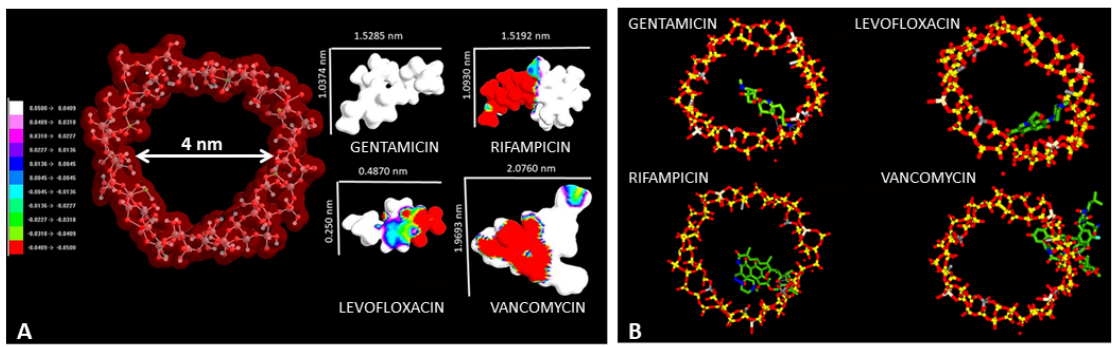

Figure 5

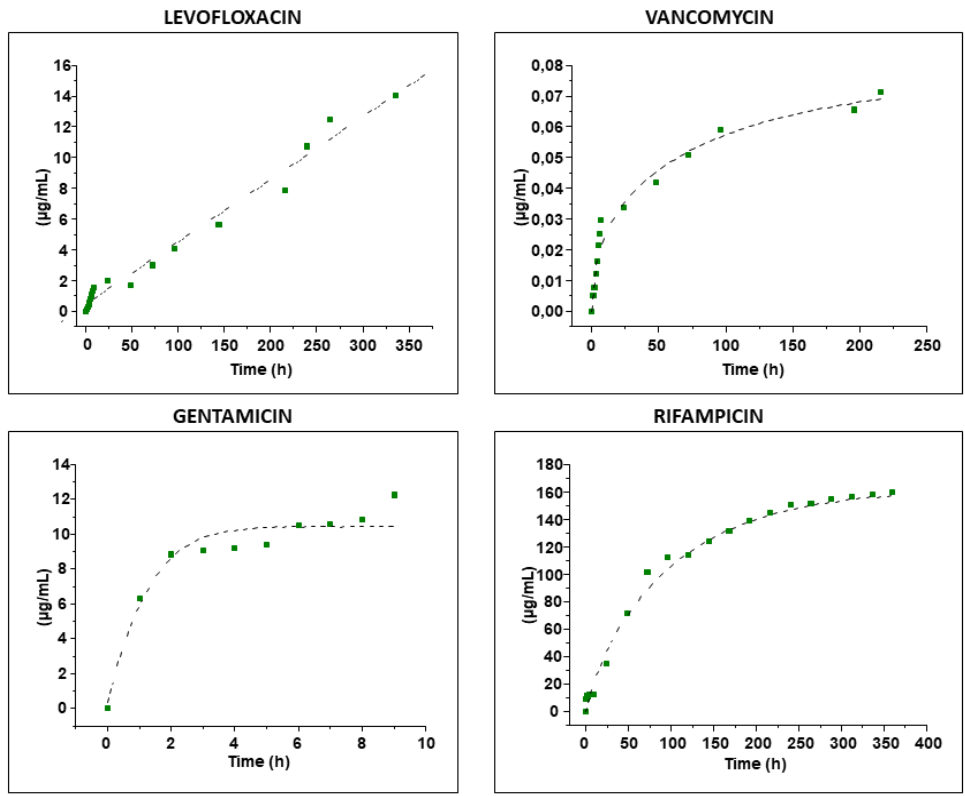

Figure 6

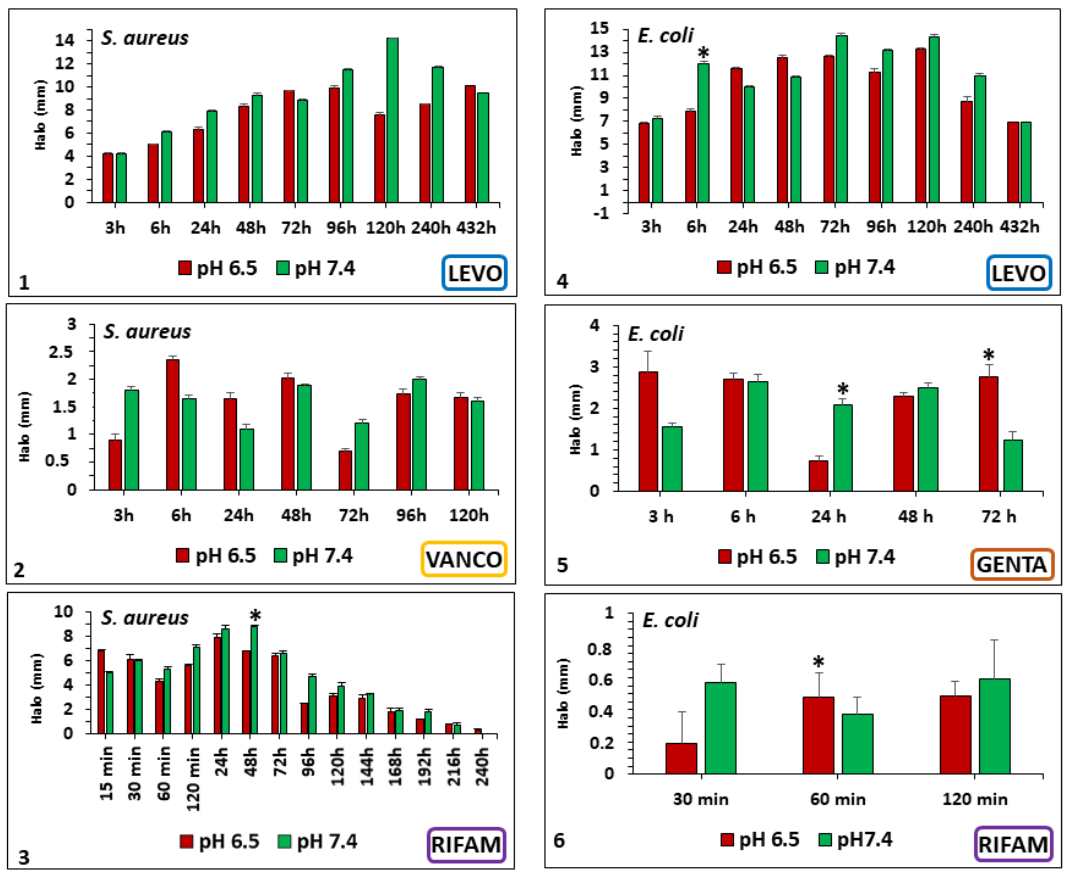

Figure 7

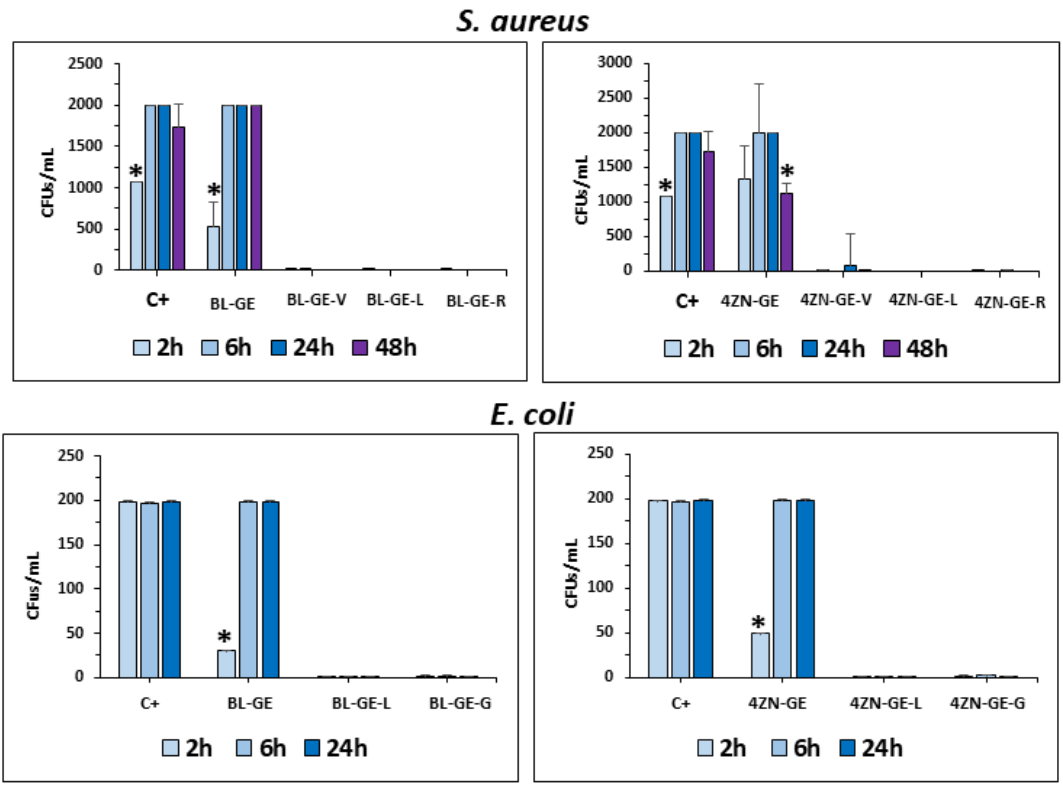

Figure 8

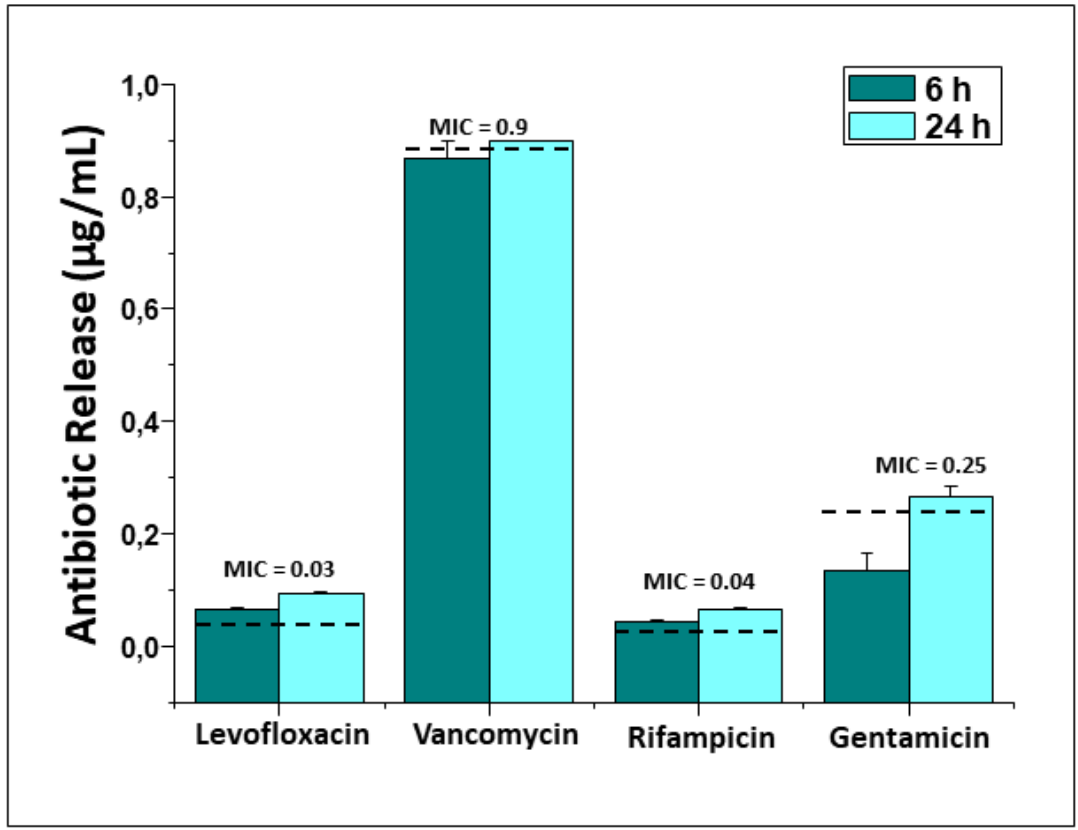

Figure 9

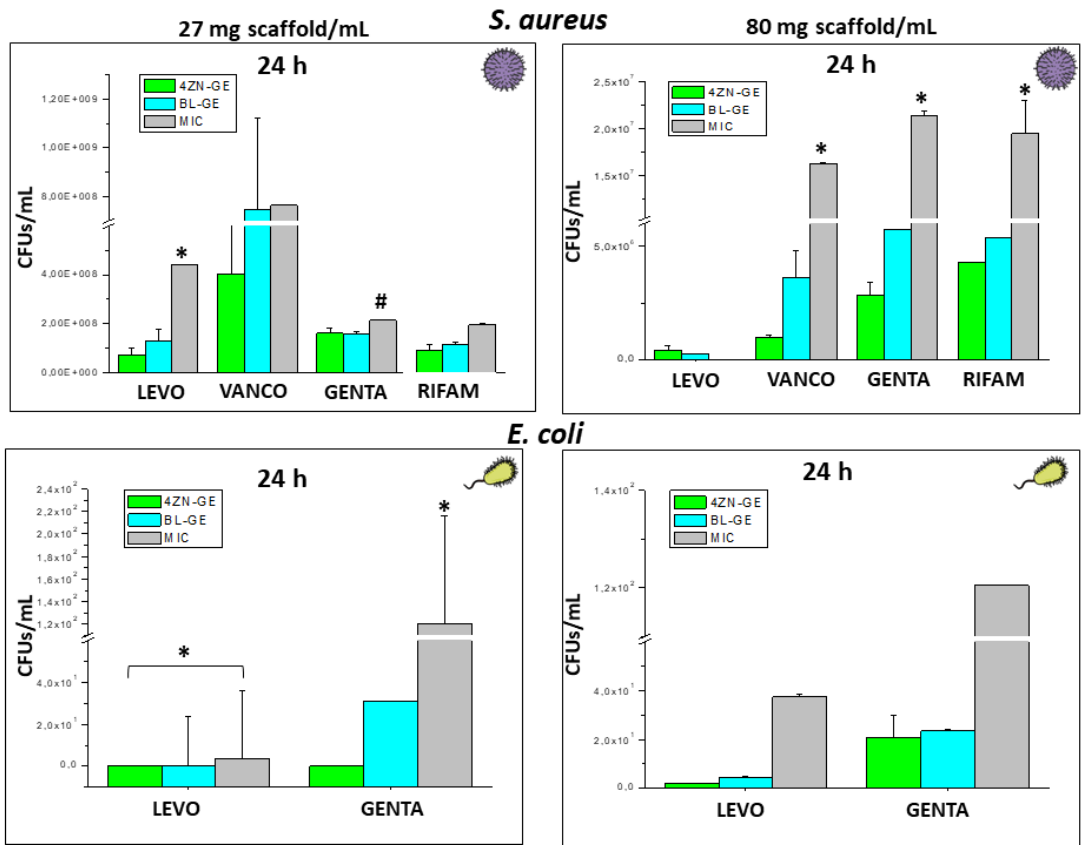

Figure 10

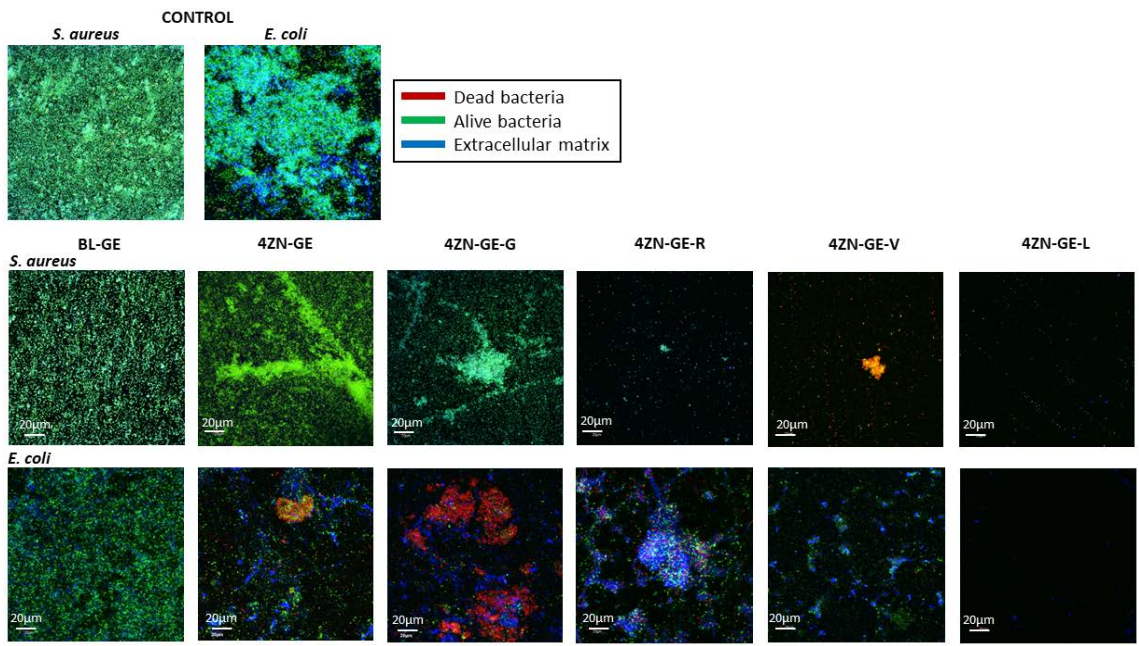

Figure 11

**Table 1**. Nitrogen adsorption porosimetry values of the scaffolds doped (4ZN) or undoped (BL) with zinc and loaded or unloaded with each antibiotic (L, V, G or R). $S_{BET}$: Brunauer-Emmett-Teller (BET) surface area; $V_p$: total pore volume obtained by the t-plot method. $D_P$: pore diameter calculated by the Barrett-Joyner-Halenda (BJH) method from the adsorption branch of isotherm. The amounts of saturated antibiotic loaded in each case form TGA were also included.

| MATERIAL | BL-GE | BL-GE-L | BL-GE-V | BL-GE-G | BL-GE-R | 4ZN-GE | 4ZN-GE-L | 4ZN-GE-V | 4ZN-GE-G | 4ZN-GE-R |
|---|---|---|---|---|---|---|---|---|---|---|
| $S_{BET}$ (m²/g) | 188 | 146.5 | 180.4 | 180.7 | 126.5 | 151.4 | 111.3 | 107.5 | 124.5 | 142.2 |
| $V_p$ (m³/g) | 0.21 | 0.15 | 0.21 | 0.21 | 0.12 | 0.14 | 0.06 | 0.11 | 0.11 | 0.13 |
| $D_P$ (nm) | 3.7 | 3.5 | 3.7 | 3.7 | 3.4 | 3.5 | 3.3 | 3.5 | 3.4 | 3.5 |
| mol drug /mg scaffold | 0 | $1.3 \times 10^{-4}$ | $1.1 \times 10^{-5}$ | $1.3 \times 10^{-5}$ | $8.4 \times 10^{-6}$ | 0 | $1.3 \times 10^{-4}$ | $1.1 \times 10^{-5}$ | $1.3 \times 10^{-5}$ | $8.4 \times 10^{-6}$ |

**Table 2**. Parameters of antibiotics kinetic release from BL-GE and 4ZN-GE scaffolds. $w_0$: initial loaded mass ($w_t$-% antibiotic loaded); A: maximum relative release; $k_1$: constant of Chapman model release rate and $K_0$ for the zero model release rate of LEVO; δ: non-ideality factor in Chapman model; R: goodness of fit.

| Sample | pH | $w_0$ (wt%) | A(%) | $K_1 (\times 10^3 \, h^{-1})$ $K_0$ LEVO | δ | R |
|---|---|---|---|---|---|---|
| 4ZN-GE-L | 7.4 | 6 | 71 | 0.05 | - | 0.99 |
| 4ZN-GE-V | 7.4 | 1 | 87 | 10.5 | 0.45 | 0.97 |
| 4ZN-GE-G | 7.4 | 2 | 99 | 8.0 | 0.27 | 0.97 |
| 4ZN-GE-R | 7.4 | 4 | 78 | 9.5 | 0.83 | 0.99 |